\documentclass[10pt,aps,prl,twocolumn,superscriptaddress,nobibnotes,nodoi,noeprint]{revtex4-2}

\usepackage{graphicx}
\usepackage{amsmath,physics,bm,float}
\usepackage{soul}
\usepackage[colorlinks=true,citecolor=blue,urlcolor=blue,linkcolor=blue]{hyperref}
\usepackage[dvipsnames]{xcolor}
\usepackage{amssymb}
\usepackage{natbib}
\usepackage{upgreek}
\usepackage{mathtools}
\usepackage{siunitx}
\usepackage[normalem]{ulem}

\usepackage[mathlines]{lineno}
\let\oldequation\equation\let\oldendequation\endequation
\renewenvironment{equation}{\linenomathNonumbers\oldequation}{\oldendequation\endlinenomath}
\let\oldalign\align\let\oldendalign\endalign
\renewenvironment{align}{\linenomathNonumbers\oldalign}{\oldendalign\endlinenomath}

\usepackage{xcolor}
\definecolor{ForestGreen}{RGB}{34,139,34}

\begin{document}
	
    \title{Active embracement enables autonomous tweezing in active star polymers}
	
	\author{Marco Musacchio}
	\email{marco.musacchio@hhu.de}
	\affiliation{
		Institut f{\"u}r Theoretische Physik II: Weiche Materie,
		Heinrich-Heine-Universit{\"a}t D{\"u}sseldorf, Universit{\"a}tsstra{\ss}e 1,
		D-40225 D{\"u}sseldorf, 
		Germany}

    \author{Davide Breoni}
	\affiliation{
    Department of Physics, Universita di Trento, Via Sommarive 14, I-38123 Trento, Italy
    }
    \affiliation{
		INFN-TIFPA, Trento Institute for Fundamental Physics and Applications, I-38123 Trento, Italy}

    \author{Iman Abdoli}
	\affiliation{
		Institut f{\"u}r Theoretische Physik II: Weiche Materie,
		Heinrich-Heine-Universit{\"a}t D{\"u}sseldorf, Universit{\"a}tsstra{\ss}e 1,
		D-40225 D{\"u}sseldorf, 
		Germany}

    \author{Luca Tubiana}
	\affiliation{
    Department of Physics, Universita di Trento, Via Sommarive 14, I-38123 Trento, Italy
    }
    \affiliation{
		INFN-TIFPA, Trento Institute for Fundamental Physics and Applications, I-38123 Trento, Italy}

	\author{Hartmut L{\"o}wen}
	\affiliation{
		Institut f{\"u}r Theoretische Physik II: Weiche Materie,
		Heinrich-Heine-Universit{\"a}t D{\"u}sseldorf, Universit{\"a}tsstra{\ss}e 1,
		D-40225 D{\"u}sseldorf, 
		Germany}

    \author{Lorenzo Caprini}
	\email{lorenzo.caprini@uniroma1.it}
	\affiliation{
		Physics department, University of Rome La Sapienza, P.le Aldo Moro 5, IT-00185 Rome, Italy}
	
	\date{\today}
        
	\begin{abstract} 
    The capacity for autonomous structural reconfiguration is a defining trait of living systems, yet it remains elusive in artificial active matter. Here, we report the discovery of \textit{active embracement}, a non-equilibrium phenomenon where active star polymers—comprising a central core and self-propelled monomeric arms -- transition from open configurations to tightly collapsed, ``hugging'' states. By combining polymer experiments using connected vibrobots with simulations, we demonstrate that internal self-propulsion fundamentally overrides the steric repulsion that keeps passive polymers dispersed. This active drive enables a suite of behaviors unattainable in equilibrium systems: individual star polymers undergo a globular-like self-collapse, multiple star polymers mutually intertwine and mutually embrace, and can spontaneously embrace and capture surrounding passive particles. Our findings reveal that active embracement is a distinct kinetic phase that allows star polymers to function as autonomous tweezers. By bridging the gap between macroscopic robotic collectives and microscopic polymer physics, this work provides a versatile blueprint for the design of smart materials capable of targeted cargo capture and self-directed assembly in complex environments.
	\end{abstract}
	
	\maketitle
    
	 \begin{figure*}[ht!]
		\includegraphics[width=0.9\linewidth]{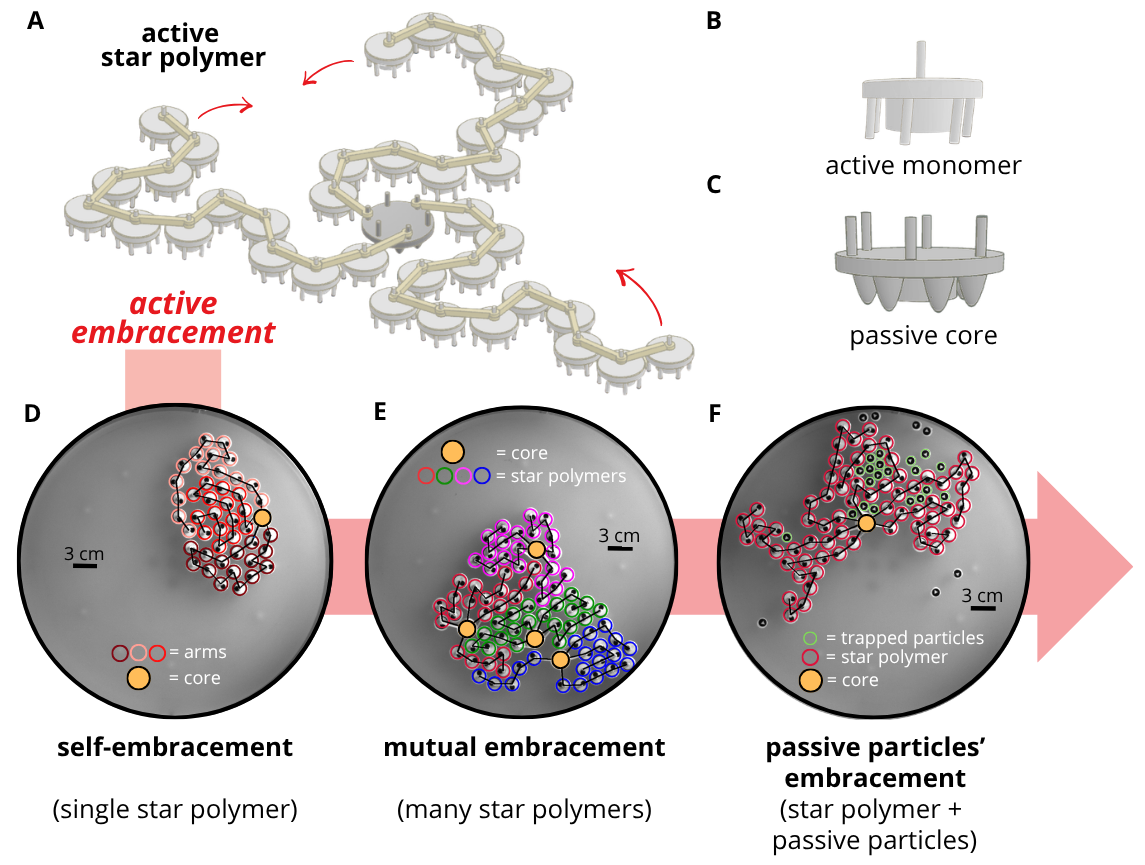}
		\caption{\textbf{Active embracement.} \textbf{A} Schematic illustration of an active star polymer composed of a passive core (large dark grey particle) and arms made of active monomers (small light grey particles).
        \textbf{B,C} Side views of the 3D-printed active monomer (\textbf{B}) and passive core (\textbf{C}).
        Red arrows indicate different manifestations of \textit{active embracement}, namely the tendency of active star polymers to ``embrace'' surrounding objects, including themselves (self-embracement), other star polymers (mutual embracement), and passive objects (embracement of passive particles).
        \textbf{D--F} Experimental snapshots showing: \textbf{D} self-embracement of a single active star polymer, with monomers belonging to different arms shown in different colors; \textbf{E} mutual embracement leading to the formation of a four-star-polymer cluster, with each star polymer colored differently; \textbf{F} an active star polymer (red monomers) capturing passive particles (gray), with captured particles highlighted in green (see \textit{Methods}). In all experimental snapshots, the passive core is colored orange, and the scale bar corresponds to $3\,\si{\centi\meter}$.}
		\label{fig:fig1}
	\end{figure*}

  \section{Introduction}

The capacity for autonomous structural reconfiguration, i.e., the ability to self-organize shape and function without external guidance, is a hallmark of living systems. Across biological scales, from cytoskeletal networks that remodel cell architecture to chromatin undergoing motor-driven reorganization~\cite{schaller2010polar,weber2015random,misteli2020self,davidson2019dna}, soft matter harnesses internally generated forces to execute folding, wrapping, and coiling~\cite{wang2020mechanism, goychuk2023polymer,caprini2025spontaneous}. These adaptive strategies enable locomotion, protection, and environmental interaction~\cite{soon2023pangolin}. Translating this level of self-directed morphing into synthetic materials remains a profound challenge: 
local force generation must be coordinated by a deformable architecture to produce global, functional shape changes.

Active matter~\cite{ramaswamy2010mechanics,bechinger2016active,Elgeti_2015,marchetti2013hydrodynamics} offers a compelling physical framework to address this challenge. In these systems, individual constituents continuously consume energy, driving the material far from thermodynamic equilibrium \cite{fodor2016far} and giving rise to emergent phenomena including spontaneous clustering \cite{cates2015motility} and pattern formation \cite{liebchen2017collective,Caprini2020Spontaneous} due to the competition between activity and interactions accounting for excluded volume. 
When these active elements are incorporated into linear polymer architectures, local propulsion or active stresses propagate along connected backbones, inducing enhanced diffusion~\cite{foglino2019non,philipps2022tangentially} and altered relaxation dynamics~\cite{winkler2017active}.
The interplay between activity and body deformability triggers conformational behaviors~\cite{winkler2020physics,eisenstecken2017internal,vatin2024conformation,martinroca2024tangentially,janzen2025active,mousavi2019active,eisenstecken2016conformational,zhu2024non, wu2025conformation, sakaue2026physics} such as snake-like motion~\cite{isele2016dynamics,sarkar2016coarse,anand2018structure}, beating~\cite{chelakkot2014flagellar,anand2019beating}, swelling~\cite{kaiser2014unusual,kaiser2015does,anand2020conformation}, and folding \cite{caprini2025spontaneous,panda2025folding,locatelli2021activity}, as well as compactification, characterized by collapsed configurations~\cite{bianco2018globulelike,duman2018collective,liu2019configuration} and oscillatory dynamics~\cite{kumar2024emergent}.
Consequently, activity provides a mechanism for programming and regulating polymer conformations, even in the absence of attractive interactions or externally imposed folding rules.

Beyond linear and ring architectures, polymer topology itself provides an additional control parameter~\cite{tubiana2024topology}. Star polymers---multiple arms tethered to a central core---exemplify how branching tunes softness, interpenetrability, and effective interactions through arm number, length, and topology~\cite{likos1998star,jusufi1999effective,watzlawek1999phase,breoni2024conformation, breoni2026mechanically}. The resulting ultrasoft, logarithmic repulsion between star centers enables their description as penetrable colloids~\cite{likos1998star,jusufi1999effective} and underpins rich phase behaviors including reentrant melting and multiple glassy states~\cite{watzlawek1999phase,mayer2009multiple}. Introducing activity into branched polymer architectures opens a route to reconfiguration mechanisms that are unavailable to passive stars or active linear chains. 
Recent simulations have shown that correlated activity can modify the size and dynamical states of star polymers, highlighting the sensitivity of branched architectures to the microscopic organization of active forces~\cite{buglakov2025polymer}. 

Yet these studies primarily concern global conformational changes such as swelling, stretching, or multistability. A qualitatively different challenge is to design active stars whose arms coordinate inward, wrap around neighboring bodies, and thereby transform conformational change into mechanical function. 
How such local active motion can be organized by a multi-arm architecture into autonomous capture remains unexplored.
In addition, autonomously targeted, task-oriented reconfiguration---such as the capture and eventual release of surrounding objects---remains a major open challenge. Most active polymers display global shape changes, including swelling, collapse, folding, or wrapping, but they do not by themselves establish a programmable route by which a single soft object can selectively engage with and manipulate its environment. This gap motivates the search for active polymer architectures that convert internal nonequilibrium motion into localized, functional interactions with external targets.

In the present work, we discover that activity at the monomer level in a star polymer (Fig.~\ref{fig:fig1}A-C) gives rise to an \textit{active embracement} mechanism. This inherently non-equilibrium phenomenon, investigated through experiments and simulations, enables star polymers to spontaneously embrace other objects, neighboring polymers, and even themselves. \textit{Active embracement} gives rise to a range of functionalities that are inaccessible to passive polymers and emerge across different levels of complexity. Individual star polymers self-embrace (Fig.~\ref{fig:fig1}D), with one arm wrapping around the rest of the polymer and driving a transition from an open configuration to a tightly collapsed state. Multiple star polymers mutually embrace (Fig.~\ref{fig:fig1}E), forming stable clusters through persistent inter-polymer contacts. Finally, star polymers autonomously capture and grasp passive particles (Fig.~\ref{fig:fig1}F), trapping them much like a spider captures its prey \cite{robinson1969predatory}. By bridging active polymer physics and robotic collectives, we demonstrate that active embracement provides a versatile design principle for engineering smart materials capable of autonomous cargo capture and self-directed assembly.

\section{Results}

\subsection{Experimental setup}

The experimental system consists of centimeter-sized granular particles fabricated by 3D printing using a proprietary photopolymer.
We designed an active star polymer (Fig.~\ref{fig:fig1}A) composed of active vibrobots \cite{antonov2025self} as arm monomers and a passive vibrobot acting as the polymer core.
Active and passive vibrobots are cylindrical particles equipped with seven legs that enable motion when the particles are placed on a vertically vibrating plate. 
Active vibrobots have all seven legs tilted at the same angle (Fig.~\ref{fig:fig1}B). The tilted legs break the translational symmetry of the body and induce a persistent self-propelled motion, which randomly reorients after a typical persistence time. In contrast, the passive core has a slightly larger diameter and is equipped with legs that are perpendicular to the vibrating plate (Fig.~\ref{fig:fig1}C). As a result, the polymer core is both rotationally and translationally symmetric and therefore exhibits passive Brownian motion.

A vertical flag mounted on each particle allows neighboring vibrobots to be connected through rigid 3D-printed linkers positioned above the particles. A star polymer consists of a passive core with $f$ arms, each composed of $N$ active monomers. The rigid linkers enforce a fixed distance between connected monomers while allowing the particles to rotate freely with respect to one another on the vibrating plate. The star polymers are placed inside a circular arena surrounded by a plastic ring that confines the system. Further details about the particle design are provided in the \textit{Methods} section.

\subsection{Simulation model}

We model the dynamics of the experimentally designed active star polymers with two-dimensional stochastic equations for their active monomers and passive core.
The passive core evolves as a passive particle, performing underdamped Brownian motion, as the particle body is translationally and rotationally symmetric without preferential directions \cite{caprini2024emergent}.
Since the active monomers have tilted legs, they propel in the direction of the tilt, which defines their polarization vector $\mathbf{n}$. As a result, a monomer can be modeled as an active Brownian particle. Consequently, each active monomer is subject to a self-propulsion force $\gamma v_0 \mathbf{n}$, which provides a constant running velocity $v_0$ directed along the particle orientation $\mathbf{n}$ \cite{antonov2025self}.

Our centimeter-sized particles exhibit inertial dynamics that endow both the particle translational velocity, $\mathbf{v}=\dot{\mathbf{x}}$, and angular velocity, $\omega=\dot{\theta}$, with finite memory times. Here, $\mathbf{x}$ denotes the particle position, while $\theta$ is the orientation angle defining the unit polarization vector $\mathbf{n}=(\cos\theta,\sin\theta)$. Furthermore, both passive and active vibrobots are subject to a Stokes-friction force,  $-\gamma \mathbf{v}$, in the center of mass dynamics, and a rotational friction force, $-\gamma_r \omega$, in the dynamics of $\theta$. Here, $\gamma$ and $\gamma_r$ are the translational and rotational friction coefficients, respectively \cite{scholz2018inertial}. These friction forces 
originate from the continuous contact friction with the substrate and can be tuned by changing the shaker's conditions \cite{caprini2024dynamical}. 
Passive and active vibrobots are exposed to a continuous injection of energy through effective white noises, induced by the random roughness of the plate and the particle design (see the \textit{Methods} section for details).

The granular monomers of a star polymer interact through potential forces, accounting for volume-exclusion effects and interaction friction forces, dissipating energy during collisions. The latter include both tangential contact and rolling friction terms.
In addition, neighboring monomers of the star polymer interact through a potential force that keeps their distance almost fixed, accounting for the effect of the rigid linkers that connect them.
Further details regarding the interactions are reported in the \textit{Methods} section.

To better assess the role of activity in our study, we have compared the results with simulations of passive star polymers, whose monomers perform Brownian motion with a vanishing self-propulsion velocity ($v_0=0$). Additionally, we have performed simulations without tangential contact and rolling friction in the dynamics of the active star polymers to shed light on the effect of these features on the phenomena experimentally observed (see \textit{Methods}).

\subsection{Self-embracement in a single active star polymer}

 \begin{figure*}[t!]
		\includegraphics[width=1\linewidth]{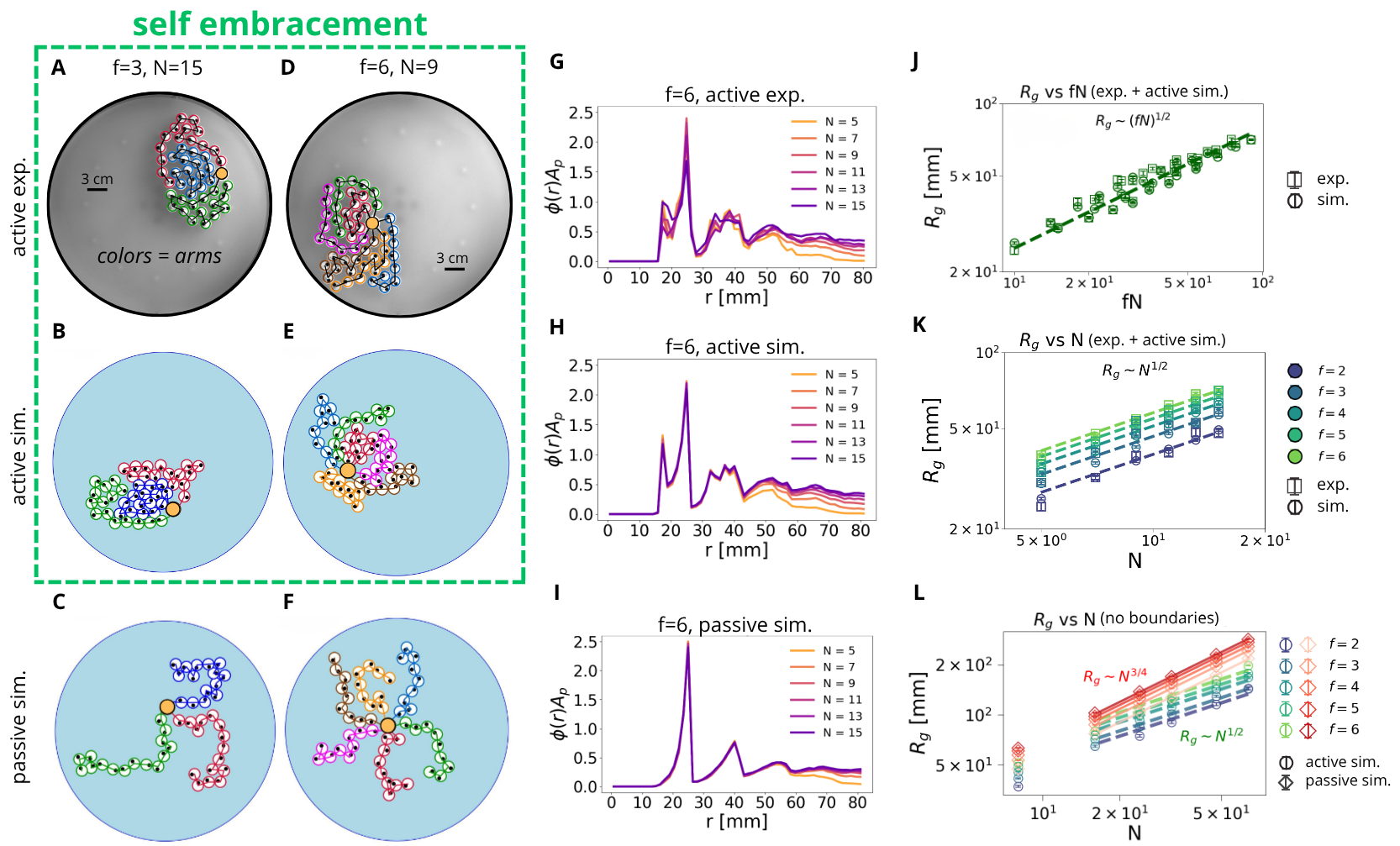}
		\caption{\textbf{Self-embracement} \textbf{A,D} Experimental snapshots of a single star-polymer with $f=3$, $N=15$ (\textbf{A}) and $f=6$, $N=9$ (\textbf{D}). \textbf{b,e} Simulation snapshots of active star-polymers for $f=3$, $N=15$ (\textbf{B}) and $f=6$, $N=9$ (\textbf{E}). \textbf{C,F} Simulations snapshots of passive star-polymers for $f=3$, $N=15$ (\textbf{C}) and $f=6$, $N=9$ (\textbf{F}). 
        In all snapshots, the passive core is colored orange, while the other colors are used to distinguish monomers belonging to different arms, and the scale bar corresponds to $3\,\si{\centi\meter}$.
        Active star polymers spontaneously adopt self-embraced configurations, whereas passive star polymers remain in open configurations. \textbf{G--I} Radial distribution functions, $\phi(r)$, of the monomer positions with respect to the passive core for a star with $f=6$ and several values of $N$ (color gradient). The distribution is multiplied by the area of a single active monomer, $A_p$, and it is shown for experiments (\textbf{G}) and simulations of active (\textbf{H}) and passive star polymers (\textbf{I}). \textbf{J} Radius of gyration $R_g$ as a function of $fN$ for both experiments (green squares) and simulations (green circles). The dashed dark green line represents the scaling $\sim(fN)^{1/2}$. \textbf{K} $R_g$ as a function of $N$ for both experiments (squares) and simulations (circles), with colors corresponding to different values of $f$ and the dashed lines representing the curve $\sim(N)^{1/2}$. \textbf{L} $R_g$ as a function of $N$ for simulations of passive (diamonds in the reddish colormap) and active (circles in the greenish colormap) star polymers, where the color gradient represents the different numbers of arms, $f$. The dashed reddish and greenish lines indicate the scalings $\sim N^{3/4}$ and $N^{1/2}$, respectively. The parameters used in the simulations to model the active monomers are reported in Table \ref{tab:act_params}, while the parameters governing the pairwise interactions are listed in Table \ref{tab:params}.}
		\label{fig:fig2} 
	\end{figure*}

We have experimentally discovered that an active granular star polymer tends to remain in a collapsed configuration, where one or more arms hug each other or the entire polymer, effectively wrapping the passive core (see Fig.~\ref{fig:fig2}A,D). 
In the steady state, one can observe that the self-hugging configuration is dominant, as the polymer rarely opens, and when it does, it collapses shortly after (see Supplementary Video 1). We refer to this phenomenon as self-embracement, being typical of active star polymers independently of the number of arms, $f$, and their number of monomers, $N$.

\begin{figure}[t!]
		\includegraphics[width=1\linewidth]{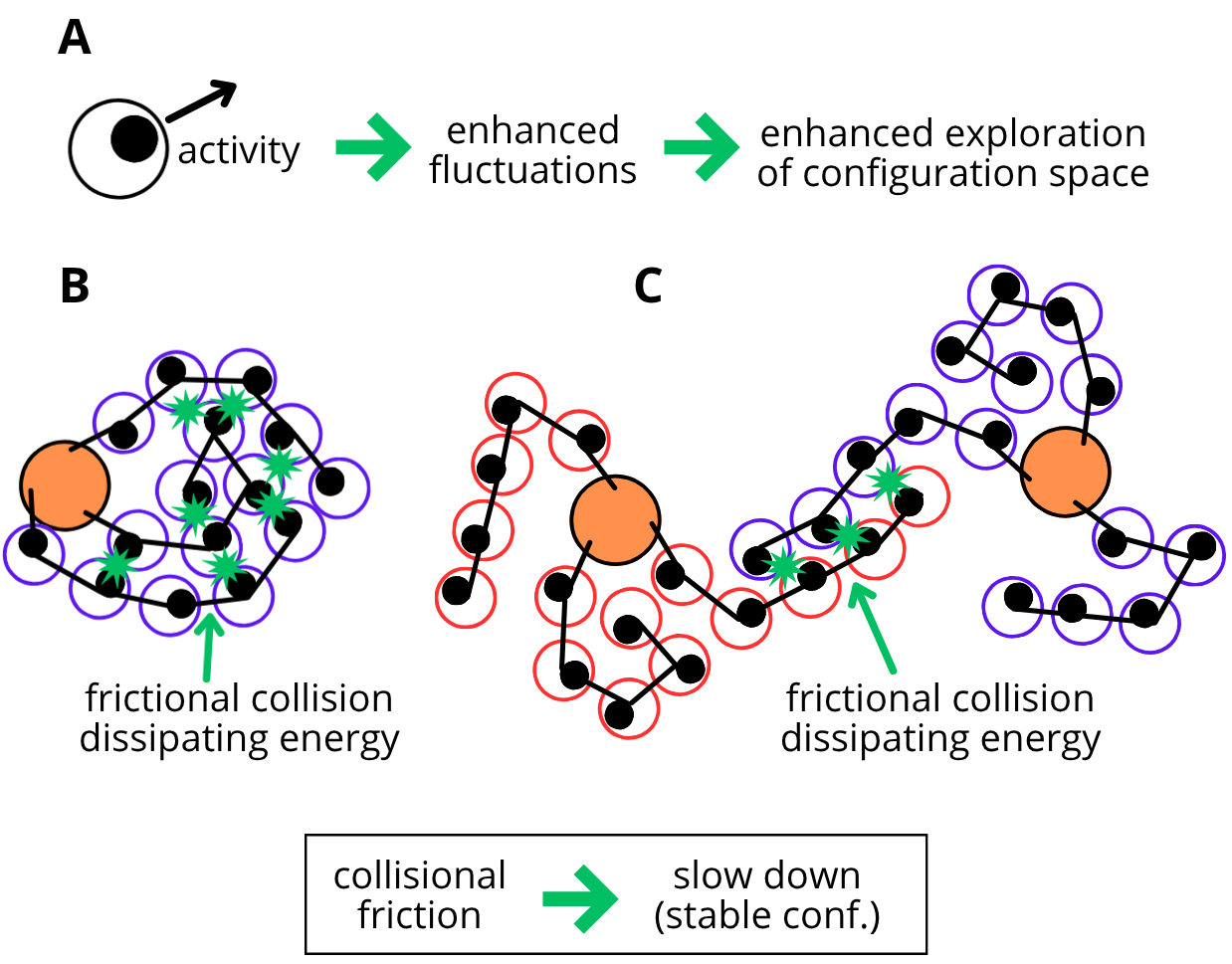}
		\caption{\textbf{\textit{Active embracement} mechanism.} 
        \textbf{A} Self-propulsion, whose direction is indicated by the black off-centered dot on each monomer, enhances fluctuations and promotes the exploration of the configuration space of the star polymer.
        \textbf{B} Illustration of a typical collapsed configuration for a single active star polymer whose active monomers are colored in blue, in which the arms are hugging each other around the passive core.
        \textbf{C} Illustration of a typical configuration involving two active star polymers, whose active monomers are colored blue and red, respectively. Configurations involving multiple monomer--monomer contacts are more stable because collisions between contacting monomers dissipate energy through collisional friction. Such contacts can occur either between different arms of the same star polymer (self-embracement) or between the arms of different polymers (mutual embracement). Passive cores are colored orange, and the monomers belonging to the same arm are connected by black lines.}
		\label{fig:fig_scheme}
	\end{figure}

The self-embracement mechanism arises from the interplay between activity and contact friction forces. Activity increases the effective temperature of each active monomer, allowing the arms to explore the surrounding space, collide with one another, and even wrap around neighboring arms (Fig.~\ref{fig:fig_scheme}A). This behavior is highly unlikely in passive star polymers, where each arm tends to avoid the others, eventually creating blobs for large arm lengths~\cite{grest2009star, de1983statistics}.
The dissipative nature of frictional collisions hinders the ability of active monomers to detach once they come into contact with surrounding particles, effectively slowing down their motion. This effect becomes more pronounced as the number of monomers in contact increases, thereby stabilizing self-embraced configurations in our active granular chains (Fig.~\ref{fig:fig_scheme}B).

To confirm this mechanism, we numerically evolve the dynamics of a single star polymer, comparing the results with and without activity for different numbers of arms and monomers per arm.
In the active case (Fig.~\ref{fig:fig2}B,E), we reproduce the experimentally observed self-embracement effect, whereas in the passive case (Fig.~\ref{fig:fig2}C,F), each arm tends to oscillate independently around the passive core without collapsing. 
Furthermore, in the absence of contact friction forces, the polymer does not show self-embracement (See Fig. \ref{fig:fig8} in the Methods), since arms continuously fluctuate and are not capable of blocking each other. These numerical studies prove the fundamental role of both dissipative collisions and activity in the observation of self-embracement.

The collapsed configuration assumed by the active star polymers can be quantified by studying the radius of gyration of the polymer, $R_g$, which measures its spatial extension, and is defined as 
$R_g=\sqrt{\left\langle
\sum_{i=1}^{N_{tot}} m_i
\left|\mathbf{x}_i-\mathbf{x}_{\mathrm{cm}}\right|^2
/\sum_{i=1}^{N_{tot}}m_i
\right\rangle}$,
where $\mathbf{x}_{\mathrm{cm}}=
\sum_{i=1}^{N_{tot}}m_i\mathbf{x}_i
/\sum_{i=1}^{N_{tot}}m_i$ denotes the center of mass of the polymer, $\mathbf{x}_i$ and $m_i$ are the particles' positions and masses, respectively, and $N_{tot}$ is the total number of particles forming the star-polymer.

The scaling behavior of the radius of gyration, $R_g$, with the number of monomers per arm, $N$, and the number of arms, $f$, provides insight into the structural properties of the star polymer. In the two-dimensional passive case, scaling arguments based on the blob theory of de Gennes and the Daoud--Cotton model predict $R_g \sim f^{1/4}N^{3/4}$ ~\cite{Daoud1982}, reflecting the tendency of excluded-volume interactions to keep the polymer arms extended. In contrast, both experiments and simulations reveal that active star polymers exhibit a scaling behavior with exponent $1/2$ with respect to both the total number of active monomers, $fN$ (Fig.~\ref{fig:fig2}J), and the number of monomers per arm, $N$, at fixed $f$ (Fig.~\ref{fig:fig2}K), implying the same scaling with $f$ when $N$ is held constant. 
The deviation from the passive scaling is not a consequence of the finite size or circular confinement of the experimental arena, as confirmed by simulations performed without confining boundaries (Fig.~\ref{fig:fig2}L). In these simulations, the expected passive scaling is recovered in the absence of activity, whereas the active case exhibits the predicted exponent of $1/2$ over a wide range of values of $f$ and $N$. The scaling law $R_g \sim (fN)^{1/2}$ observed for active star polymers is consistent with the collapsed configurations found in both experiments and simulations. Indeed, this is the same scaling expected for the radius of a compact two-dimensional object composed of $fN$ particles, supporting the interpretation of self-embraced star polymers as dynamically self-collapsed structures.

The compact structure adopted by the active star polymers, both in experiments and in simulations, is also well characterized by the radial distribution function of particles around the passive core. The radial distribution function is defined as $\phi(r) = \frac{1}{2\pi r\mathrm{d}r} \left\langle \sum_{i}^{Nf} \delta\left(r - |\mathbf{x}_i - \mathbf{x}_{\mathrm{core}}|\right) \right\rangle$, where $\mathbf{x}_i$ is the particle's position, $\mathbf{x}_{\mathrm{core}}$ is the position of the core, $Nf$ is the total number of monomers in the system, and $\mathrm{d}r$ is the radial bin width. Multiplying this quantity by the area of a single active monomer, $A_p$, yields a dimensionless function which, in the active case (Fig.~\ref{fig:fig2}G,H), exhibits several peaks. The first peak, both in experiments and simulations, occurs at a distance corresponding to the typical separation between the passive core and the attached active monomers. The subsequent structure of the distribution shows two distinct contributions: peaks located at integer multiples of the rigid link length, arising from the underlying chain connectivity, and additional features at smaller distances. The latter originate from the self-embracement effect, which promotes a local reorganization of the monomers and drives them to collapse towards the passive core.
In contrast, simulations of a passive star polymer (Fig.~\ref{fig:fig2}I) show a markedly different density profile, characterized by fewer peaks. In the absence of any tendency to collapse around the passive core, the distribution is dominated solely by the geometric constraints imposed by the rigid links. As a result, only peaks located at integer multiples of the link length are observed, while any additional structure associated with local collapse is absent. This indicates that the polymer arms remain effectively extended and are constrained primarily by their connectivity rather than by any active reconfiguration.

The self-embracement, namely the tendency of an active star polymer to hug its own arms and wrap around its core, represents the simplest manifestation of a more general phenomenon that we term \textit{active embracement}. As we show in the following sections, \textit{active embracement} is not limited to self-interactions: active star polymers can also embrace other star polymers or even isolated passive particles.

\subsection{Mutual embracement of multiple active star polymers}

\begin{figure*}[t!]
		\includegraphics[width=1\linewidth]{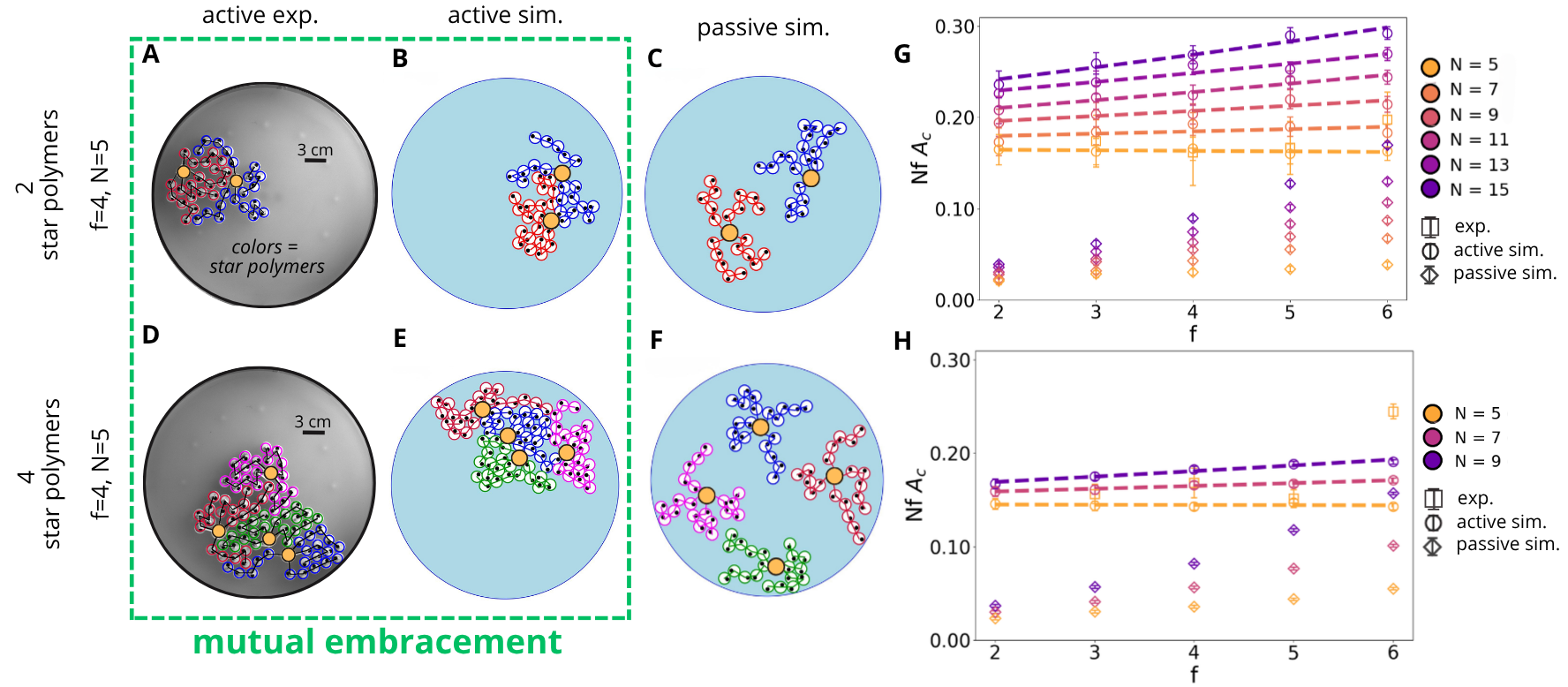}
		\caption{\textbf{Mutual embracement.} \textbf{A,D} Experimental snapshots of two (\textbf{A}) and four (\textbf{D}) star polymers with number of arms $f=4$ and number of monomers per arm $N=5$.  \textbf{B,C,E,F} Simulation snapshots of two (\textbf{B}) and four (\textbf{E}) active star polymers and two (\textbf{C}) and four (\textbf{F}) passive star polymers, with $f=4$ and $N=5$ to match experiments. 
        The passive core of each star polymer is colored orange, while all monomers belonging to the same polymer have the same color. In all snapshots, the scale bar corresponds to $3\,\si{\centi\meter}$.
        In the active case, different star polymers cluster and remain in contact through mutual embracement, whereas in the passive case they effectively repel each other.
        \textbf{G,H} Contact area, $A_c$, multiplied by the total number of particles forming a single star polymer, $Nf$, as a function of $f$. $A_c$ is calculated for experiments (squares), simulations with active (circles) and passive (diamonds) star polymers, for two (\textbf{G}) and four (\textbf{H}) interacting star polymers. In the experiments just $N=5$ is explored. The color gradient represents different values of $N$. The parameters used in the simulations to model the active monomers are reported in Table \ref{tab:act_params}, while the parameters governing the pairwise interactions are listed in Table \ref{tab:params}.}
		\label{fig:fig3}
	\end{figure*}

When two or more active star polymers interact, their arms do not simply tend to wrap around their own cores, but also to embrace neighboring polymers (see Supplementary Video~2). We refer to this collective behavior as mutual embracement. It can be regarded as a manifestation of the general \textit{active embracement} mechanism introduced above, since it originates from the same underlying physical process. When the arms of different polymers mutually embrace, several monomers come into contact simultaneously, and the resulting contact and rolling frictions effectively slow down their relative motion (Fig.~\ref{fig:fig_scheme}C), favoring configurations in which neighboring polymers remain in close proximity (Fig.~\ref{fig:fig3}A,D).
This embracement mechanism promotes the aggregation of active star polymers, which therefore tend to form clusters. Remarkably, regardless of the number of polymers in the arena, the number of arms $f$ or the number of monomers per arm $N$, experiments show that active star polymers spend most of the time in close proximity to one another, with their arms wrapped around neighboring polymers. The same aggregation process is reproduced in simulations of active star polymers (Fig.~\ref{fig:fig3}B,E). Passive polymers instead tend to remain separated, behaving as if they were subject to effective repulsive interactions (Fig.~\ref{fig:fig3}C,F).

The extent of mutual embracement can be quantified by measuring the contact area between two or more star polymers as a function of the number of arms, $f$. The contact area, $A_c$, is defined as
$
A_c =
\frac{1}{\sum_{\alpha < \beta} (Nf)_{\alpha}(Nf)_{\beta}}
\sum_{\alpha < \beta}
\sum_{i \in \alpha}
\sum_{j \in \beta}
\Theta\left(d-\left|\mathbf{x}_i-\mathbf{x}_j\right|\right),
$
where $\alpha$ and $\beta$ denote different polymers, $(Nf)_\alpha$ represents the number of active particles in polymer $\alpha$, $\mathbf{x}_i$ denotes the position of particle $i$, and $d$ corresponds to the cutoff distance below which particles belonging to different polymers are considered to be in contact. Accordingly, $A_c$ measures the fraction of inter-polymer particle pairs that are in contact. Multiplication by the total number of monomers in a star polymer, $fN$, yields a quantity that approaches unity when, on average, each particle of one polymer is in contact with a particle belonging to another polymer. As such, $A_c$ provides a direct measure of the extent of mutual embracement, capturing the degree of interpenetration between neighboring polymers beyond what can be inferred from the average core-to-core distance alone.

Figure~\ref{fig:fig3}G,H reveals a clear quantitative distinction between active and passive systems. The contact area measured for active star polymers (squares for experiments and open circles for simulations) is consistently larger than that of passive star polymers (diamonds), both for systems comprising two and four interacting polymers. These higher values provide direct evidence of activity-induced mutual embracement. By increasing the extent of inter-polymer contact, mutual embracement stabilizes configurations in which neighboring polymer cores remain in close proximity, thereby promoting clustering. This behavior stands in sharp contrast to the passive case, where star polymers behave according to effective repulsive interactions.

\subsection{Passive particles' embracement by a single active star polymer}

\begin{figure*}[t!]
		\includegraphics[width=1\linewidth]{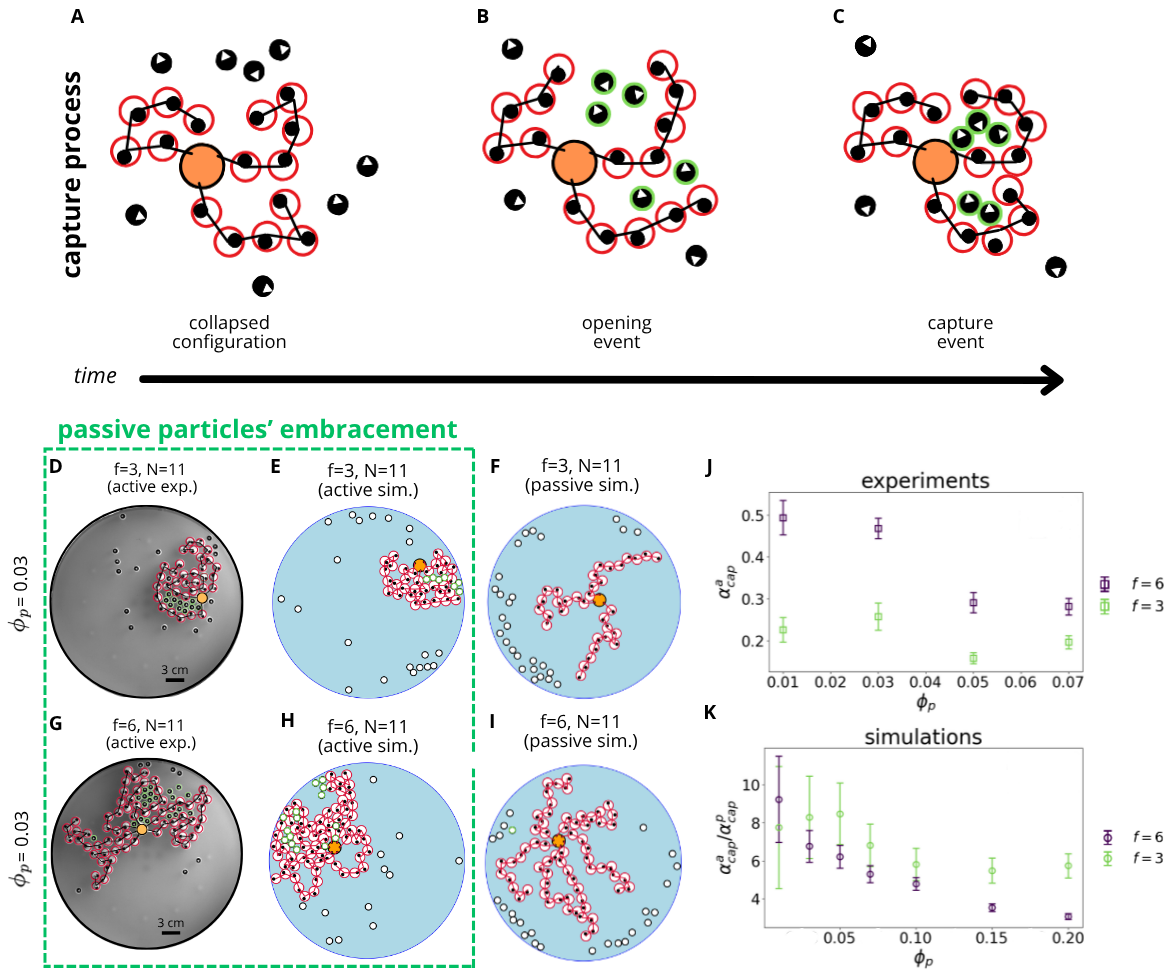}
		\caption{\textbf{Passive particles' embracement.} \textbf{A-C} Illustration of the capture process of passive particles by an active star polymer. The passive core of the star polymer is colored orange, arm monomers are colored red, while captured passive particles are highlighted in green (see \textit{Methods}). \textbf{D,G} Experimental snapshots of a single star polymer with number of arms $f=3$ (\textbf{D}) and $f=6$ (\textbf{G}).  In the experimental snapshots, the scale bar corresponds to $3\,\si{\centi\meter}$. \textbf{E,H} Simulation snapshots of an active star polymer with $f=3$ (\textbf{E}) and $f=6$ (\textbf{H}) and the corresponding passive passive star polymer with $f=3$ (\textbf{F}) and $f=6$ (\textbf{I}). In the panels \textbf{D}-\textbf{I}, the number of monomers per arm is fixed at $N=11$, the packing fraction of the passive particles is $\phi_p=0.03$, and we use the same color convention as in the schematic illustrations. \textbf{J} Fraction of passive particles captured in the experiments, $\alpha_{cap}^{a}$, as a function of $\phi_p$ for $f=6$ (purple squares) and $f=3$ (green squares) at $N=11$. \textbf{K} Ratio between the fraction of passive particles captured by the active star polymer, $\alpha_{cap}^{a}$, and that captured by the passive star polymer, $\alpha_{cap}^{p}$, obtained from simulations, as a function of $\phi_p$. Results for the star polymer with $f=6$ are shown as purple circles, while those for the star polymer with $f=3$ are shown as green circles. The parameters used in the simulations to model active monomers and passive core are reported in Table~\ref{tab:act_params} and Table~\ref{tab:pass_params}, while the parameters governing the pairwise interactions are listed in Table~\ref{tab:params}.}
		\label{fig:fig4}
	\end{figure*}

When an active star polymer is immersed in a bath of passive particles, \textit{active embracement}, namely its tendency to wrap the arms around surrounding objects, enables it to capture nearby particles (see Fig.~\ref{fig:fig4}D,G). We refer to this process as embracement of passive particles. To experimentally characterize this behavior, we systematically varied the passive particle packing fraction, $\phi_p$, first using a star polymer with $N=11$ and $f=3$, and then repeating the same experiments with a star polymer having $N=11$ and $f=6$. As illustrated in Fig.~\ref{fig:fig4}A-C, the polymer continuously wrapped and unwrapped its arms, transiently enclosing nearby passive particles and thereby capturing and transporting them (see Supplementary Video~3). Numerical simulations of active star polymers reproduce the same qualitative behavior (see Fig.~\ref{fig:fig4}E,H). By contrast, passive star polymers (see Fig.~\ref{fig:fig4}F,I) are unable to capture nearby particles, as steric interactions prevent the arms from enclosing them (see Supplementary Video~4).

To quantify the ability of an active star polymer to capture passive particles, we define the fraction of captured particles, $\alpha_{\mathrm{cap}}$, as the fraction of passive particles enclosed within the instantaneous polymer configuration. At each time step, the positions of all monomers composing the star polymer are used as input points to construct the convex hull. This is defined as the smallest convex polygon containing all monomer positions, obtained by connecting the outermost monomers of the polymer configuration. This polygon provides a time-dependent region that approximately identifies the area enclosed by the star polymer arms. A passive particle is considered captured if its position lies inside this polygon. The captured fraction is therefore calculated as
$\alpha_{\mathrm{cap}}(t)=N_{\mathrm{cap}}(t)/N_p$,
where $N_{\mathrm{cap}}(t)$ is the number of passive particles located inside the convex hull at time $t$, and $N_p$ is the total number of passive particles in the system. By evaluating $\alpha_{\mathrm{cap}}(t)$ over the whole trajectory, we obtain the temporal evolution of the capture process. The average captured fraction is then calculated as the time average of $\alpha_{\mathrm{cap}}(t)$. In Fig.~\ref{fig:fig4}J, we report the experimental measurements of the fraction of passive particles captured by the active star polymer, $\alpha_{\mathrm{cap}}^{a}$ (where the superscript $a$ denotes the active case), as a function of the passive particle packing fraction, $\phi_p$. The purple squares correspond to a star polymer with $N=11$ and $f=6$: at low values of $\phi_p$, $\alpha_{\mathrm{cap}}^{a}$ approaches values close to $0.5$, indicating that, on average, nearly half of the passive particles are captured by the star polymer. For the star polymer with $N=11$ and $f=3$ (green circles), the captured fraction is lower, suggesting that increasing the number of arms enhances the particle capture efficiency. Nevertheless, even in this case, $\alpha_{\mathrm{cap}}^{a}$ remains significantly above zero, demonstrating the persistence of the  \textit{active embracement} effect.

In Fig.~\ref{fig:fig4}K, we report the simulation results for the ratio between the fraction of passive particles captured by the active star polymer, $\alpha_{\mathrm{cap}}^{a}$ (where the superscript $a$ denotes the active case), and the fraction captured by the passive counterpart, $\alpha_{\mathrm{cap}}^{p}$ (where the superscript $p$ denotes the passive case). These simulations were performed over a wider range of passive particle packing fractions compared to the experimental study. At low particle densities, the active star polymer is able to capture nearly one order of magnitude more particles than the passive polymer. This ratio decreases with increasing $\phi_p$, as the number of available free particles becomes larger and the relative advantage of the active mechanism is reduced. Nevertheless, even at the highest packing fractions investigated, the active star polymer captures approximately four times more particles than its passive counterpart. 
These results reveal \textit{active embracement} as a robust mechanism enabling star polymers to transiently capture and transport passive particles. The qualitative agreement between experiments and simulations confirms that the active rearrangement of the polymer arms is the key ingredient responsible for this enhanced particle uptake, which remains significantly larger than in the passive counterpart across the investigated range of particle densities.

\section{Discussion}

In this work, we present the first experimental realization of an active star polymer, in which multiple active arms are attached to a passive core. We show that the interplay between persistent self-propulsion and contact friction gives rise to a distinct non-equilibrium mechanism, which we term \textit{active embracement}. This phenomenon is characterized by the spontaneous tendency of active star polymers to wrap their flexible arms around surrounding objects. At the single-particle level, the arms embrace one another, driving the polymer into a compact self-collapsed state through a process of self-embracement. In suspensions of multiple active star polymers, the same mechanism promotes aggregation, with neighboring polymers becoming wrapped together into long-lived clusters through mutual embracement.

Beyond the specific experimental realization presented here, our results suggest that \textit{active embracement} constitutes a general mechanism by which activity, body architecture, and friction cooperate to generate effective mechanical interactions. Unlike conventional self-organization driven by attractive forces, the observed wrapping behavior emerges dynamically from the persistent propulsion of flexible appendages in contact with neighboring objects. In this sense, activity does not simply enhance transport or diffusion, but endows structured active bodies with an autonomous grasping capability that is absent in their passive counterparts. More broadly, our findings demonstrate that body topology should be regarded as a fundamental design parameter of active matter, alongside propulsion and interactions, for controlling emergent mechanical functionality.

An intriguing perspective opened by this work concerns the collective behavior of dense suspensions of active star polymers, so far mainly investigated in the case of active linear polymers~\cite{day2024morphological, patil2023ultrafast, deblais2020phase, ozkan2021collective,vahid2025collective}. The continuous formation and disruption of arm-mediated contacts effectively act as a form of active and reversible crosslinking, giving rise to transient entanglement networks and dynamically evolving mechanical connections. Such active topological constraints may produce collective phenomena with no counterpart in conventional active particles or passive polymeric materials, including non-equilibrium viscoelasticity, yielding, and stress-relaxation dynamics~\cite{breoni2025giant}. We therefore envision active star polymers as a unique experimental platform at the interface between active matter and polymer physics, where the interplay between activity, topology, and mechanical constraints remains largely unexplored.

Finally, the \textit{active embracement} mechanism enables active star polymers to autonomously capture, transport, and release passive objects by wrapping their flexible arms around the cargo. This behavior is qualitatively reminiscent of grasping strategies employed by biological organisms such as spiders~\cite{robinson1969predatory}, octopuses~\cite{gutfreund1998patterns}, or potentially worm–star aggregates~\cite{hodgkin2013two}, although it arises entirely from physical self-organization rather than biological control. Beyond demonstrating autonomous grasping at the macroscale, our results suggest new design principles, based on structured active polymers, for cargo transport, active carriers, and soft microrobotic systems capable of grasping objects with different shapes~\cite{jones2021bubble, becker2022active}. More generally, exploiting the coupling between activity, flexible architecture, and friction may provide a route toward active materials capable of performing increasingly sophisticated mechanical tasks without centralized control.

\section{Methods}

\subsection{Experimental details}

\begin{figure}[t!]
		\includegraphics[width=1\linewidth]{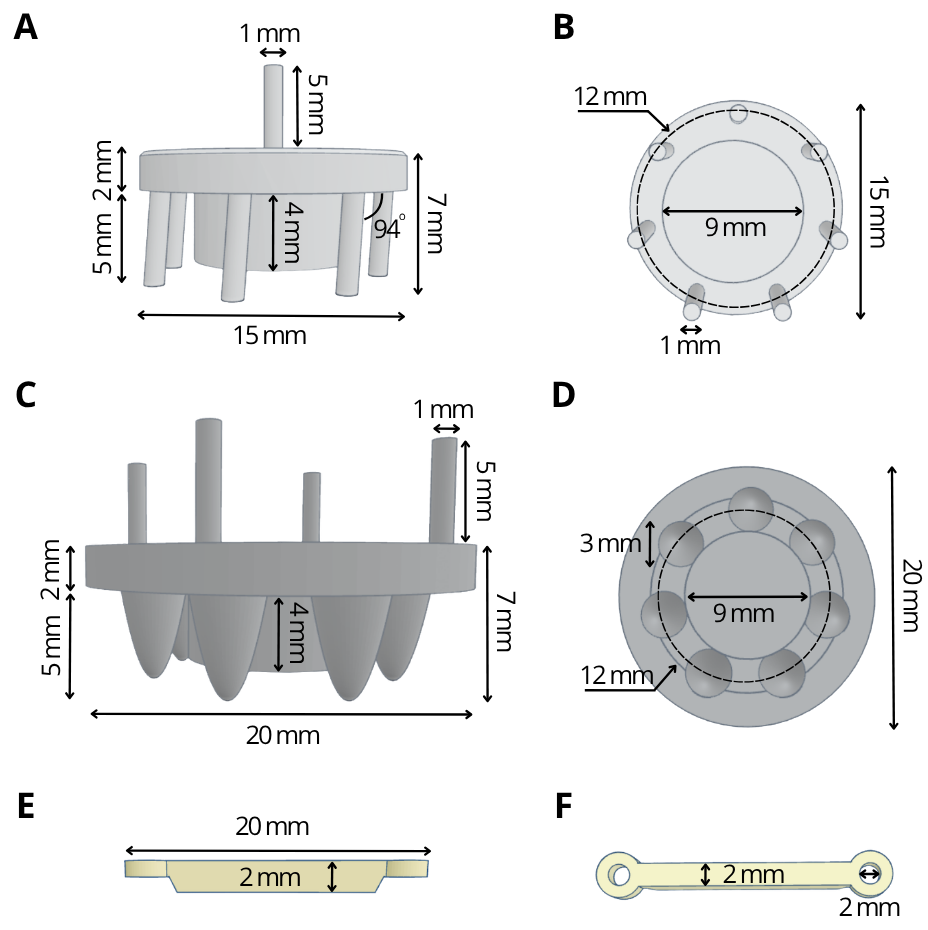}
		\caption{
        \textbf{3D model of the active star polymer.}
        \textbf{A-D} Side and bottom views of the 3D-printed active granular particle used as an active monomer (\textbf{A},\textbf{B}) and the passive granular particle used as the star polymer core (\textbf{C},\textbf{D}). We report the heights of the particle components, the details for the legs, including tilting angle and diameter, and the details of the cylinders forming the body of the particles and the vertical flags. \textbf{E,F} Side and bottom views of the 3D-printed rigid link, showing the length and the width.}
		\label{fig:fig5}
	\end{figure}

\noindent
\textbf{Particle design.}
Active and passive granular particles forming, respectively, the arms of the star polymer and the passive core to which the arms are attached, are manufactured using a proprietary photopolymer in a stereolithographic 3D printer. 
Each particle has a cylindrical body consisting of two concentric cylinders~\cite{scholz2018inertial}.
The upper cylinder (the particle cap) has a height of $2\,\si{\milli\meter}$ and a diameter $\sigma = 15\,\si{\milli\meter}$ for the active particles (Fig.~\ref{fig:fig5}~A, B) and $\sigma_c = 20\,\si{\milli\meter}$ for the passive particles (Fig.~\ref{fig:fig5}~C, D), while the lower one (the particle core) has a height of $4\,\si{\milli\meter}$ and a diameter of $9\,\si{\milli\meter}$ for both the active and the passive particles. Each active particle touches the plate via seven cylindrical legs with a radius of $0.5\,\si{\milli\meter}$ and a height of $5\,\si{\milli\meter}$. These legs are attached to the cap and are tilted in the same direction at an angle of $4^\circ$. Each particle has a mass of $0.86 \pm 0.01\,\si{g}$ and features a vertical cylindrical flag mounted at the center of the upper cylinder. The flag has a height of $5\,\si{\milli\meter}$ and a radius of $0.5\,\si{\milli\meter}$. The passive particles are also in contact with the vibrating plate via seven legs; however, in this case the legs are not tilted and have a conical shape, with a base radius of $3\,\si{\milli\meter}$ and a flattened tip at ground level. The passive particles have a mass of $1.53 \pm 0.01,\,\si{g}$ and can be equipped with four, five, or six vertical cylindrical flags, each with a height of $5\,\si{\milli\meter}$ and a radius of $0.5\,\si{\milli\meter}$. The flags are mounted on the upper cylinder at a radial distance of $8.5\,\si{\milli\meter}$ from its center and are evenly spaced along the circumference.

A white label sticker with a black border is placed on top of each particle to assist the tracking algorithm, while a black spot is included to denote the particle's orientation.

The particles are connected by rigid links of length $20\,\si{\milli\meter}$, featuring holes at their ends separated by $17\,\si{\milli\meter}$ that accommodate the vertical pegs on top of the particles. The holes at the ends of each link are aligned along one edge, ensuring precise alignment when two links are joined through the vertical flag of a particle.

\vskip10pt
\noindent
\textbf{Experimental setup and particle motion.}
The active star-polymer consists of a passive core from which $f$ arms of $N$ active monomers each extend. The experiments explore values of $f$ in the range $2 \le f \le 6$ and $N$ in the range $5 \le N \le 15$. The particles are connected by rigid links that allow free rotation of the active monomers. The arena in which the particles are placed consists of an acrylic plate with a diameter $D = 300\,\si{\milli\meter}$ that undergoes vertical vibrations. Plate oscillations are induced by an electromagnetic shaker driven by a signal from a conventional function generator and amplified accordingly. As confirmed in a previous study with a similar setup~\cite{caprini2024emergent}, the plate oscillations are spatially homogeneous and transfer the same amount of energy to each active granular particle. Each particle performs vertical jumps whose amplitude increases with the shaker amplitude, while their period is set by the shaker frequency. 
Regarding the active monomers, the tilted legs break the translational symmetry of the particle, leading to asymmetric jumps. This asymmetry results in directed motion of the particle on the plate upon release of elastic energy.
Since the motion in the vertical direction, for both active and passive particles, is small compared to the horizontal one, the dynamics of an active granular system can be effectively described as quasi-two-dimensional. Finally, imperfections of the plate and particles, together with residual vertical motion, generate an additional effective translational noise that, alongside the self-propulsion speed, contributes to driving the dynamics. For all the experiments performed, we keep the shaker's frequency constant at 100 \si{Hz}.

\vskip10pt
\noindent
\textbf{Data acquisition.}
 Data are recorded by using a high-speed camera placed above the setup, which captures 20 images per second and is characterized by a spatial resolution of 3.22 \si{px/mm}. 
 By using a tracking algorithm, we extract particle positions while the orientations are calculated from the relative position of the black spot compared to the particle center of mass.
 In every recording images, positions and orientations are calculated with sub-pixel precision, using conventional image processing techniques.\\

 All the experiments presented below have a duration of approximately forty minutes. In all the analyses performed, we neglected the initial fifty seconds in order to avoid introducing a bias due to the initial conditions.

\begin{figure*}[t!]
		\includegraphics[width=18cm]{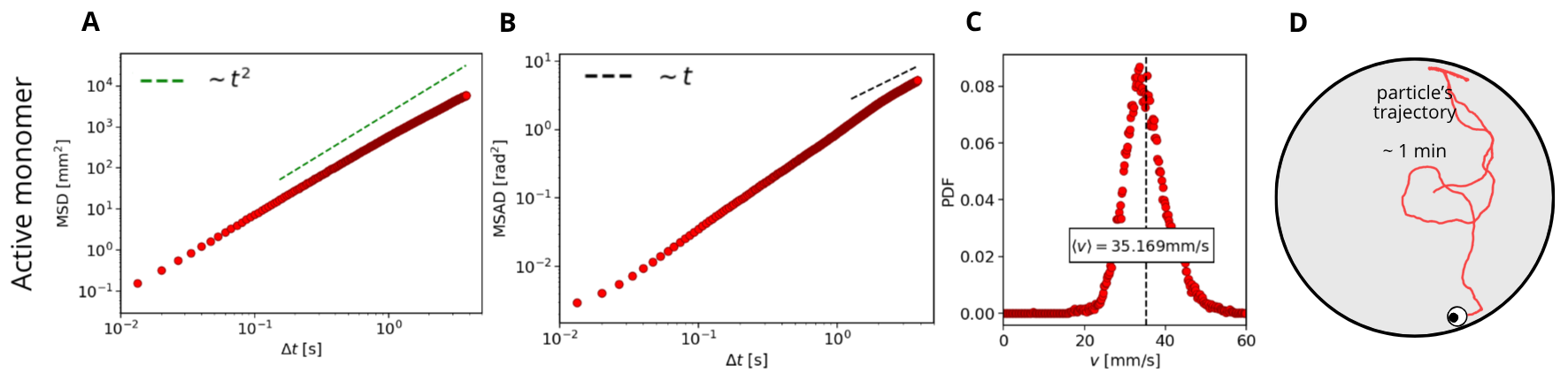}
		\caption{\textbf{Active monomer dynamics.} \textbf{A,B} Plots showing the mean square displacement (MSD) and the mean angular square displacement (MSAD), respectively, for an active monomer. The green and black dashed lines represent the $t^2$ and $t$ scaling behaviors, respectively, at sufficiently long times. \textbf{C} Plot showing the probability density function of the active monomer velocity. The black dashed line indicates the mean velocity of the distribution, $\langle v \rangle = 35.169\,\mathrm{mm\,s^{-1}}$. \textbf{D} Representative trajectory (red line) of the active monomer recorded over approximately one minute.}
		\label{fig:fig6}
	\end{figure*}

\begin{figure*}[t!]
		\includegraphics[width=18cm]{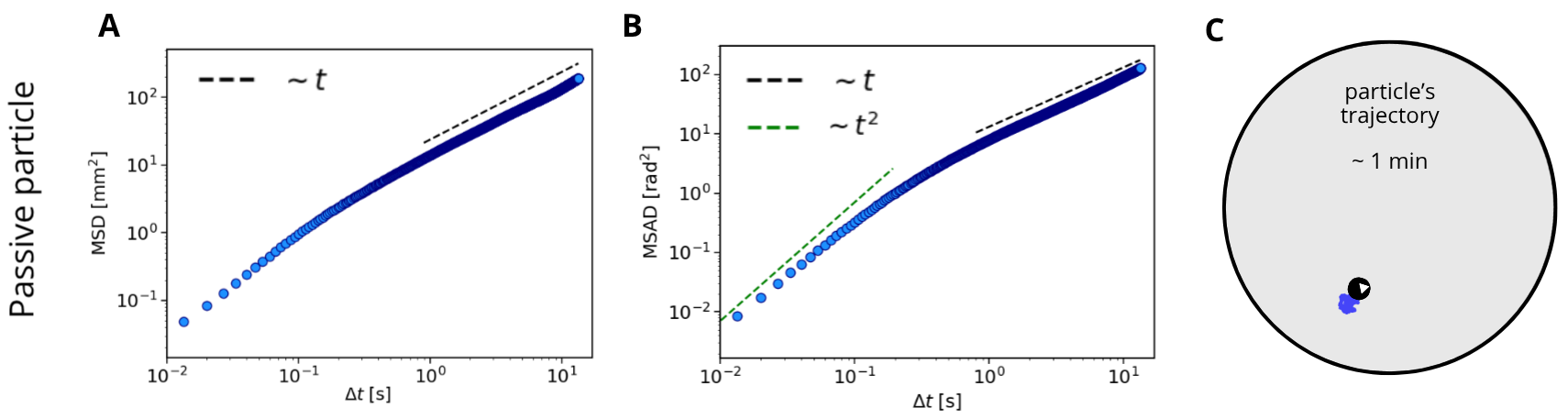}
		\caption{\textbf{Passive particle dynamics.} \textbf{A,B} Plots showing the mean square displacement (MSD) and the mean angular square displacement (MSAD), respectively, for a passive particle. The green and black dashed lines represent the $t^2$ and $t$ scaling behaviors, respectively. \textbf{C} Representative trajectory (blue line) of a passive particle recorded over approximately one minute}
		\label{fig:fig7}
	\end{figure*}

\vskip10pt
\noindent
\textbf{Parameter extraction} --
The parameters of the interaction potentials are chosen such that $\sigma$ corresponds to the nominal diameter of an active monomer. The intrinsic parameters of the active monomers are instead determined by fitting the translational velocity distribution, the mean-square displacement, and the angular mean-square displacement (Fig.~\ref{fig:fig6}A--C) over multiple single particles trajectories (look at Fig.~\ref{fig:fig6}D for a typical trajectory). We find that, under the shaker conditions considered, the active monomers are characterized by a persistence time of $1/D_r = 1.25\,\mathrm{s}$, a translational diffusion time $\sigma^2/D=2\cdot 10^{-3} s$ and an average speed of $\langle v \rangle = 35\,\mathrm{mm/s}$. The same procedure is adopted to extract the intrinsic parameters of the passive particles used in the capture experiments. By fitting the mean-square displacement, and the angular mean-square displacement (Fig.~\ref{fig:fig7}A,B) over multiple single particle trajectories (look at Fig.~\ref{fig:fig7}C for a typical trajectory), the resulting parameters are a rotational diffusion time of $1/D_r = 0.23\,\mathrm{s}$, a rotational inertial time of $J/\gamma_r = 0.1\,\mathrm{s}$ and a translational diffusion time $\sigma^2/D=4.5\cdot 10^{-3} s$. The remaining simulation parameters are carefully selected to reproduce the experimental observations as accurately as possible.

\subsection{Simulation details}
\label{sec:sims}

\subsubsection{Equation of motion of an active star polymer}

An active star polymer has a central core and $f$ arms consisting of $N$ monomers each. All the particles' dynamics are simulated through stochastic equations of motion using LAMMPS~\cite{thompson_lammps_2022}: the active monomers evolve as inertial active Brownian particles (ABPs) subject to translational and rotational Stokes friction~\cite{caprini2022role}, while the passive core evolves as an inertial passive particle.

\vskip6pt
\noindent
\textbf{Equations of motion of the active monomer} --
Each active monomer is described by a stochastic equation of motion for its translational velocity $\textbf{v}_i=\dot{\textbf{x}}_i$, and angular velocity $\omega_i=\dot{\phi}_i$, where $\textbf{x}_i$ and $\phi_i$ represent the position and the orientational angle of the particle. 
The dynamics reads
\begin{eqnarray}
m\dot{\textbf{v}}_i &=& -\gamma\textbf{v}_i+\gamma\sqrt{2D_t}\boldsymbol{\xi}_i+\gamma v_0\textbf{p}_i + \textbf{F}^e_i
\label{eq:activemonomer_trasl}\\
j\dot{\omega}_i &=& -\gamma_r\omega_i+\gamma_r\sqrt{2D_r}\xi^r_i + \mathcal{T}^e_i
\label{eq:activemonomer_orient}\,.
\end{eqnarray}
where $i=1,.., fN$, since $f$ is the number of arm and $N$ the number of monomers per arm.
Here, $\boldsymbol{\xi}_i$ and $\xi^r_i$ are Gaussian white noises with unit variance and zero average. The coefficients $\gamma$ and $\gamma_r$ represent the translational and rotational friction coefficients, while $D_t$ and $D_r$ are the translational and rotational diffusion constants. 
Each active monomer is subject to an active force $\gamma v_0 \textbf{p}_i$ which induces a constant self-propulsion velocity $v_0$ along the orientation vector $\textbf{p}_i=(\cos{\theta_i}, \sin{\theta_i})$. The terms $\textbf{F}^e_i$ and $\mathcal{T}^e_i$ are the total external forces and torques acting on the particles, accounting for volume exclusion, tangential and rolling contact frictions (discussed below).

The model parameters in Eqs.~\eqref{eq:activemonomer_trasl} and \eqref{eq:activemonomer_orient} of a single active monomer are set as in experiments. We have chosen $\gamma=100m\tau^{-1}$, where $\tau=1s$, $\gamma_r=100J\tau^{-1}$ $D_r=0.8\tau^{-1}$. In the case of active star simulations, we set the arm beads to have $v_0=2.3\sigma/\tau$, $D_t=0.01\sigma^2/\tau$.  The simulations have a timestep size of $\Delta t=10^{-4}\tau$ and run for a total of $10^9$ timesteps.

\begin{table}[h!]
\centering
\begin{tabular}{|c|c|}
\hline
\textbf{Parameter} & \textbf{Value} \\ \hline
$v_0$ & 2.3 $\sigma/\tau$ \\ \hline
$D_t$ & $0.01\sigma^2/\tau$ \\ \hline
$D_r$ & $0.8\tau^{-1}$ \\ \hline
$\gamma$ & $100m/\tau$ \\ \hline
$\gamma_r$ & $100J_p/\tau$ \\ \hline
\end{tabular}
\caption{Active particle coefficients, chosen to match experiments.}
\label{tab:act_params}
\end{table}

As mentioned in the \textit{Results} section, we have performed simulations with passive star polymers where each monomer behaves as a passive particle. In this case, we set the active speed to zero ($v_0=0$), and their diffusion coefficient to $D_t \to D_t^l=v_0^2/(2D_r)\simeq3.403\sigma^2/\tau$. The value of $D_t^l$ was chosen to give passive particles the same long time diffusion coefficient as active ones, making the comparison between active and passive stars more consistent. 

\vskip6pt
\noindent
\textbf{Equations of motion of the passive core} --
The passive core is described by an underdamped equation of motion for its velocity $\mathbf{V}=\dot{X}$ and its angular velocity $\Omega=\dot{\Theta}$:
\begin{eqnarray}
M\dot{\textbf{V}}&=&-\gamma\textbf{V}+\gamma\sqrt{2D_t}\boldsymbol{\xi} + \textbf{F}^e \\
\label{eq:activemonomer_trasl_p}
J\dot{\Omega}&=&-\gamma_r\Omega+\gamma_r\sqrt{2D_r}\xi^r + \mathcal{T}^e \,,
\label{eq:activemonomer_orient_p}
\end{eqnarray}
where $\boldsymbol{\xi}$ and $\xi^r$ are white noise terms with zero average and unit variance.
The equations for the translational and angular velocities are not coupled since the passive core has no self-propulsion velocity.
The dynamical parameters $\gamma$, $\gamma_r$, $D_t$, and $D_r$ are chosen equal to those for active monomers since they mainly depend on the particle material and the shaker's conditions.
As in the case of active monomers, the terms $\textbf{F}^e$ and $\mathcal{T}^e$ are the total external forces and torques acting on the particles.

\vskip6pt
\noindent
\textbf{Structure of the active star} --
Simulations are performed as close to the experimental setup as possible: the stars are composed of $f$ arms of $N$ beads each, connected by one of their ends to $f$ pins. These pins in turn form, together with the central bead, a rigid body. We focus on star polymers with $2\leq f\leq6$ and $5\leq N\leq 15$. Setting the diameter of an arm bead $\sigma=15mm$, we have that the central bead diameter is $\sigma_c=4/3\sigma$, the arm bond size at rest is $b_0=17/15\sigma$, and the pins are positioned at a fixed distance from the central bead $r_p=8.5/15\sigma$. The stars can move within a circular confinement of diameter $20\sigma$. The mass of the arm beads is set as $m$, the central core has mass $M=\left(\frac{\sigma_c}{\sigma}\right)^3m$, since the particles are considered to be circular, and the pins have negligible mass. The moment of inertia $J$ of the various particles is calculated as $J=1/10m\sigma^2$, still considering a spherical shape.

\subsubsection{Interactions among monomers}

\vskip6pt
\noindent
Since our particles are granular, they are not only subject to volume exclusion interactions but also to dissipative collisions, resulting in static and dynamic tangential and rolling friction forces.
So the center of mass of the monomer $i$ is subject to a total force $\mathbf{F}_i$ given by the sum of several pairwise contributions:
\begin{equation}
\label{eq:total_force}
    \mathbf{F}_i = \sum_{j\neq i} (\mathbf{F}^p_{ij} + \mathbf{F}^n_{ij} + \mathbf{F}^t_{ij})
\end{equation}
where $\mathbf{F}^p_{ij}$ is the force contribution due to a conservative potential, which is used to model the fixed distance between neighboring monomers; the term $\mathbf{F}^n_{ij}$ is calculated through the Hertz model and represents the interactions directed along the distance connecting the centers of two particles. This force includes standard repulsion between colliding particles (elastic contribution) accounting for volume exclusion effects and a damping term (dissipative contribution) related to the normal component of the relative velocity between colliding particles.
Finally, the term $\mathbf{F}^t_{ij}$ is modeled through a modified Mindlin model~\cite{mindlin_compliance_1949}. This term contains both an elastic contribution and a dissipative contribution related to the tangential component of the relative velocity between colliding particles.

Additionally, the monomer $i$ is subject to a total torque $\mathcal{T}_i$, which is given by the sum of two pairwise contributions
\begin{equation}
    \mathcal{T}_i = \sum_{j\neq i} (\mathcal{T}^t_{ij} + \tau^r_{ij}) \,,
\end{equation}
with the sum running over all the other particles. The term $\mathcal{T}^t_{ij}$ is the torque generated by the tangential force $\mathbf{F}^t_{ij}$, previously introduced, while $\tau^r_{ij}$ is the torque calculated from a spring-dashpot-slider (sds) model~\cite{luding_cohesive_2008} typically employed to account for the rolling friction.
We remark that both $\mathbf{F}^n_{ij}$ and $\mathbf{F}^p_{ij}$ do not generate any torque, being directed along the direction connecting the centers of the two interacting particles.

We remark that the same force contributions reported in Eq.~\eqref{eq:total_force} are used to model the interactions between the particle $i$ and the circular boundary.
Below, we describe in details all the force and torque terms employed in the numerical study and report the model parameters.

\vskip6pt
\noindent
\textbf{Bond interactions} --
The force $\mathbf{F}_{ij}^{p}$ accounting for the fixed distance between neighboring monomers can be derived from a pairwise potential, such that $\mathbf{F}^p_{ij} = - \nabla_i U_b(|\mathbf{x}_j -\mathbf{x}_i|)$. Here, $U_b$ is chosen as a FENE (Finite Extensible Nonlinear Elastic) potential~\cite{grest_molecular_1986} $U_b(r)=U_r(r)+U_a(r)$, where $r$ is the distance between neighboring monomers, and $U_r(r)$ and $U_a(r)$ are the repulsive and attractive components, respectively:
\begin{equation}
\label{Ur}
U_r(r)=
\begin{cases}
4\epsilon \left[\left(\frac{b_0}{r}\right)^{12}-\left(\frac{b_0}{r}\right)^6\right]+\epsilon, &r\leq r_{c},\\
0 ,&r> r_{c},
\end{cases}
\end{equation} 
where $r_c=2^{1/6}b_0$ is the cutoff distance and $\epsilon=10m\sigma^2/\tau^2$ is the bond stiffness;
\begin{equation}
\label{Us}
U_a(r)=
\begin{cases}
-\frac{KR_0^2}{2}\text{ln} \left[1-\left(\frac{r}{R_0}\right)^2\right],&r\leq R_0,\\
\infty, &r>R_0 \,.
\end{cases}
\end{equation}
Here, $R_0=1.5 b_0$ is the maximum bond length and $K=30\epsilon/\sigma^2$ is the strength of the spring.
The latter parameter is chosen large to ensure that the distance between neighboring monomers remain almost constant to the value $b_0$, which represents the nominal length of a rigid linker.

\vskip6pt
\noindent
\textbf{Normal interactions: The Hertz model} --
The repulsive force $\mathbf{F}^n_{ij}$ is calculated through the Hertz model and determined by two contributions: an elastic one, modeling the repulsion due to volume exclusion, $\textbf{F}^{n,e}_{ij}$, and a dissipative term, $\textbf{F}^{n,d}_{ij}$, slowing down the particles, such that
\begin{equation}
    \textbf{F}^n_{ij}=\textbf{F}^{n,e}_{ij}+\textbf{F}^{n,d}_{ij} \,.
\end{equation}
 Both forces are directed along the direction connecting the centers of two interacting particles, described by the unit vector $\textbf{n}_{ij}=\textbf{r}_{ij}/|\textbf{r}_{ij}|$, with $\textbf{r}_{ij}=\textbf{x}_i-\textbf{x}_j$.

The first contribution $\textbf{F}^{n,e}_{ij}$ is modeled as an elastic contribution and is given by
\begin{equation}
    \textbf{F}^{n,e}_{ij}
    =k_n\sqrt{R_{eff}\delta^3_{ij}}\textbf{n}_{ij} \,,
\end{equation}
where $R_{eff}=\frac{\sigma_i\sigma_j}{2(\sigma_i+\sigma_j)}$ is the effective radius, $\delta_{ij}=(\sigma_i+\sigma_j)/2-|\textbf{r}_{ij}|$ represents the particle overlap and $k_n$ is the spring constant. 

The second contribution is a damping force depending on the velocities of the interacting particles and reads
\begin{equation}
    \textbf{F}^{n,d}_{ij}=-\eta_n (\textbf{n}_{ij}\otimes\textbf{n}_{ij}) \cdot (\textbf{v}_i-\textbf{v}_j)\,.
\end{equation}
This expression projects the relative velocity $(\textbf{v}_i-\textbf{v}_j)$ along the direction $\textbf{n}_{ij}$ connecting the centers of the two interacting particles. The constant $\eta_n$ represents the normal damping coefficient, related to the restitution coefficient $e$ with the equation
\begin{equation}
    \eta_n=-2\sqrt{\frac{5}{6}}\frac{\log(e)}{\sqrt{\pi^2+(\log(e))^2}}\sqrt{\frac{3}{2}k_{n}am_{eff}} \,,
\end{equation}
where $m_{eff}=m_im_j/(m_i+m_j)$ is an effective mass and the term $a$ is the radius of the contact region, defined as
\begin{equation}
\label{eq:contact_region}
    a=\sqrt{R_{eff}\delta_{ij}}
\end{equation}
depending on the effective radius and the particle overlap.

\vskip6pt
\noindent
\textbf{Tangential interactions: The Mindlin model} --
Granular particles are typically subject to an additional force directed tangentially compared to the line connecting the centers of the two particles responsible for energy dissipation.
This force is modeled through the Mindlin model and reads
\begin{equation}
    \textbf{F}^t_{ij}=-\min(\mu_t|\textbf{F}^n_{ij}|,|\textbf{F}^{t,e}_{ij}+\textbf{F}^{t,d}_{ij}|)\textbf{t}_{ij},
\end{equation}
where $\textbf{t}_{ij}$ is the vector normal to $\mathbf{n}_{ij}$.
This force is given by the sum of a tangential elastic contribution $\textbf{F}^{t,e}_{ij}$ and a dissipative contribution $\textbf{F}^{t,d}_{ij}$ but is calculated as the minimum between this force and $\mu_t|\textbf{F}^n_{ij}|$, where $\mu_t$ is the static friction coefficient. The choice between these two contributions accounts for the static friction that characterizes granular particles.

The dissipative contribution depends on the tangential component of the relative velocity between two interacting particles
and tends to align the tangential component of the velocities of particles $i$ and $j$: indeed, it has the following expression
\begin{equation}
    \textbf{F}^{t,d}_{ij}=-\eta_t \mathbf{v}^t_{ij}
\end{equation}
where $\eta_t$ represents the tangential damping coefficient and the term $\mathbf{v}^t_{ij}$ represents the relative velocity projected onto the direction transverse to $\mathbf{n}_{ij}$.
Therefore, its expression is given by
\begin{equation}
    \mathbf{v}^t_{ij} = (I - \mathbf{n}_{ij}\otimes \mathbf{n}_{ij}) \cdot (\mathbf{v}_i - \mathbf{v}_j) + \sigma (\omega_i + \omega_j)\mathbf{z}\times\mathbf{n}_{ij}
\end{equation}
where $\mathbf{z}$ is a unit vector orthogonal to the plane of motion.
The projection operator $(I - \mathbf{n}_{ij}\otimes \mathbf{n}_{ij})$ selects only the vector component orthogonal to $\mathbf{n}_{ij}$, while the other term, depending on $\omega_j + \omega_i$, accounts for the fact that our particles have a finite size and rotate with an angular velocity.

The elastic tangential contribution is calculated incrementally, using the following expression:
\begin{equation}
   \textbf{F}^{t,e}_{ij}=-k_t\int_{t_0}^ta(t')\textbf{v}^t_{ij}(t')\text{d}t',
\end{equation}
involving the relative tangential velocity and the radius of the contact region defined in Eq. \eqref{eq:contact_region}.
Here, $k_t$ is the tangential stiffness coefficient and $t_0$ defines the time of first contact between the particles.
If contact decreases over time, i.e. $a(t)<a(t-\Delta t)$, the tangential elastic force $\textbf{F}^{t,e}_{ij}$ is rescaled~\cite{thornton_investigation_2013}:
\begin{equation}
    \textbf{F}^{t,e}_{ij}(t)=\textbf{F}^{t,e}(t-\Delta t)\frac{a(t)}{a(t-\Delta t)} \,.
\end{equation}
We note that also $\textbf{F}^{t,e}_{ij}$ is directed along $t_{ij}$ being proportional to $\textbf{v}^t_{ij}$. 

Since $\mathbf{F}^t_{ij}$ is directed along $\mathbf{t}_{ij}$, it generates a torque affecting the dynamics of the angular velocity, given by:
\begin{equation}
    \mathcal{T}^t_{ij}= R_{eff} \mathbf{n}_{ij} \times \mathbf{F}_{ij}^t
\end{equation}
where $R_{eff}$ is the effective radius previously defined.

\vskip6pt
\noindent
\textbf{The rolling friction} --
Finally, we include an additional friction term, the rolling friction, which does not affect the center of mass but only induces a torque. This term is modeled via a 2D spring-dashpot-slider model, which introduces a pseudo-force $\textbf{F}^r_{ij}$ that does not affect the total forces on the interacting particles, but is defined as a means to induce an effective torque on them, given by
\begin{eqnarray}
    \tau^r_{ij}&=&R_{eff}\textbf{n}_{ij}\times\textbf{F}^r_{ij}\\
    \tau^r_{ij}&=&-\tau^r_{ji}\,,
\end{eqnarray}
where $R_{eff}$ is the effective radius previously introduced.
The pseudo-force is defined as follows
\begin{equation}
    \textbf{F}^r_{ij}=\min(\mu_r|\textbf{F}^n_{ij}|,|k_r\boldsymbol{\zeta}_{ij}-\eta_r\textbf{v}^r_{ij}|)\textbf{k},
\end{equation}
where $\textbf{v}^r_{ij}=-R_{eff}(\omega_i-\omega_j)\textbf{z}\times\textbf{n}$ is the relative rolling velocity and $\textbf{z}$ represents a unit vector pointing orthogonally compared to the plane of motion. This force is again oriented along $\mathbf{t}_{ij}$, since the unit vector $\mathbf{k}$ is defined as $\textbf{k}=\textbf{v}^r_{ij}/|\textbf{v}^r_{ij}|$. The coefficient $k_r$ denotes the rolling elastic stiffness while $\eta_r$ represents the rolling damping coefficient. 
Such a force is given by the minimum between $\mu_r |\textbf{F}^n_{ij}|$
and $|k_r\boldsymbol{\zeta}_{ij}-\eta_r\textbf{v}^r_{ij}|$, where $\boldsymbol{\zeta}_{ij}$ denotes the rolling displacement which is computed incrementally as
\begin{equation}
    \boldsymbol{\zeta}_{ij}(t)=\int_{t_0}^t\textbf{v}^r_{ij}(t')\text{d}t'.
\end{equation}
We remark that the minimum function accounts for the static rolling friction: indeed, it selects the minimal force between the normal force and the sum between rolling displacement and dissipation. The parameters that we chose for our simulations are listed in Table~\ref{tab:params}.
\begin{table}[h!]
\centering
\begin{tabular}{|c|c|}
\hline
\textbf{Parameter} & \textbf{Value} \\ \hline
$k_n$ & $10^5m/\tau^2$ \\ \hline
$k_t$ & $10^6m/\tau^2$ \\ \hline
$k_t$ (wall) & $10^5m/\tau^2$ \\ \hline
$k_r$ & $10^5m/\tau^2$ \\ \hline
$\eta_n$ & $50\tau^{-1}$ \\ \hline
$\eta_t$ & $.1\tau^{-1}$ \\ \hline
$\eta_r$ & $50\tau^{-1}$ \\ \hline
$\mu_t$ & $10$ \\ \hline
$\mu_t$ (wall) & $1.5$ \\ \hline
$\mu_r$ & $1$ \\ \hline
\end{tabular}
\caption{Pairwise interaction parameters. Those with the (wall) qualifier refer to particle-wall interactions. If no parameter with (wall) is present, it means that the same was used both for particle-particle and particle-wall interactions.}
\label{tab:params}
\end{table}

\begin{figure*}[t!]
		\includegraphics[width=18cm]{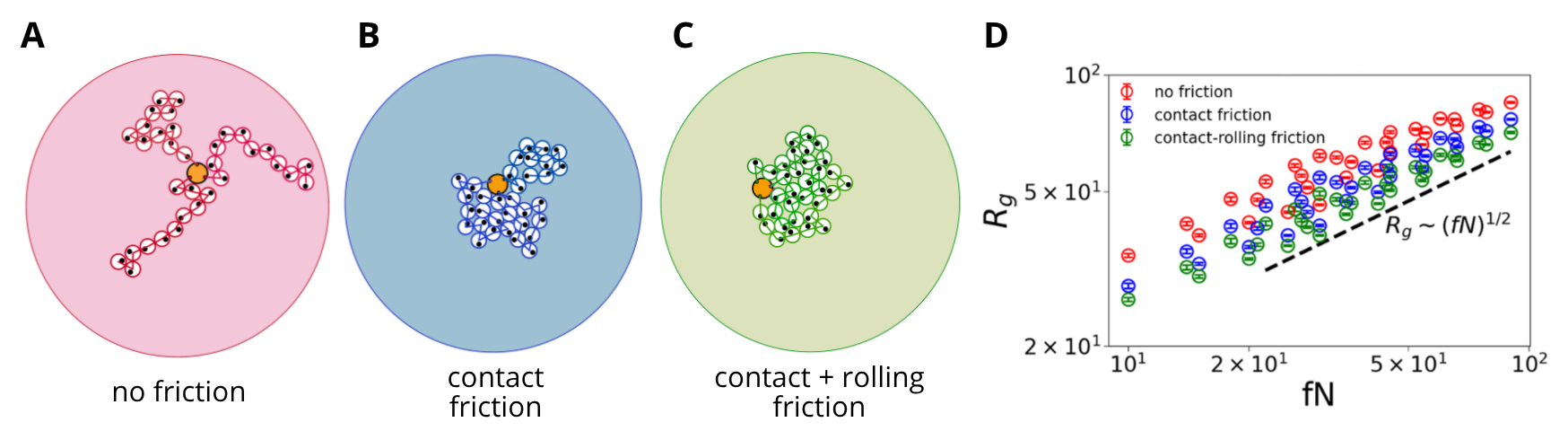}
		\caption{\textbf{Effects of pairwise friction on active star polymers.} \textbf{A-C} Simulation snapshots of active star polymers ($f=3$, $N=11$) with no pairwise friction \textbf{A}, with contact friction \textbf{B}, and with both contact and rolling friction \textbf{C}. \textbf{d} Gyration radius of active polymers in different friction regimes as a function of $f\times N$. For more information, see the \textit{Interactions between Monomers} section in the \textit{Methods}.}
		\label{fig:fig8}
	\end{figure*}

In Fig.~\ref{fig:fig8} we show the impact that the inclusion of pairwise friction has on active star polymers. Both from snapshots (Fig.~\ref{fig:fig8}A-C) and the gyration radius (Fig.~\ref{fig:fig8}D), we deduce that the signature globular configurations observed in experiments (Fig.~\ref{fig:fig1}D) are better reproduced by a system with both contact and rolling friction.

\vskip6pt
\noindent
\textbf{Capture of passive particles:}
Passive particles have a diameter $\sigma_p=2/3\sigma$, and hence a mass $m_p=\left(\frac{\sigma_p}{\sigma}\right)^3m$ and a momentum of inertia $J_p=\frac{1}{10}m\sigma_p^2$. They are modeled as passive inertial Brownian particles. Their pairwise interactions are modeled in the same way as those of the star monomers, with the same coefficients.
These interactions are the same between passive and active particles, passive and passive particles, and passive particles and walls. The number of passive particles considered is in the range $9\leq N_p\leq 180$. The parameters used for the passive particles are listed in Table~\ref{tab:pass_params}.
\begin{table}[h!]
\centering
\begin{tabular}{|c|c|}
\hline
\textbf{Parameter} & \textbf{Value} \\ \hline
$D_t^p$ & $0.015\sigma^2/\tau$ \\ \hline
$D_r^p$ & $4.4\tau^{-1}$ \\ \hline
$\gamma^p$ & $30m/\tau$ \\ \hline
$\gamma_r^p$ & $9.5J_p/\tau$ \\ \hline
\end{tabular}
\caption{Passive particle coefficients, chosen to match experiments.}
\label{tab:pass_params}
\end{table}
At each time step, the convex hull of the star polymer is constructed from the monomer positions, defining the smallest convex polygon enclosing the polymer. A passive particle is considered captured and colored green in Fig.~\ref{fig:fig1} and Fig.~\ref{fig:fig4} if its position lies within this polygon.

\section{Video description}

\noindent
\textbf{Supplementary Video 1:} Three panels are shown, corresponding to the experiment and to the simulations of active and passive star polymers, respectively, each featuring a single star polymer with three arms and fifteen monomers per arm. The active star polymer (both in the experiment and in the active simulation) spends a significant fraction of time in collapsed configurations (self-embracement), in which the arms, each displayed in a different color, embrace one another and tend to wrap around the passive core (orange full circle). In contrast, in the passive case, this embracing tendency is absent, and the arms instead fluctuate without forming persistent mutual contacts. \\

\noindent
\textbf{Supplementary Video 2:} Three separate panels are presented, showing the experiment and the simulations of active and passive star polymers, respectively. In all cases, four star polymers are shown ($f=4, N=5$), each in a different color. In both the experiment and the active simulations, the polymers tend to intertwine and form compact agglomerates (mutual embracement), a behavior that is absent in the passive case, where the star polymers instead repel each other due to steric interactions. \\

\noindent
\textbf{Supplementary Video 3:} Experimental realization of a single star polymer ($f=3$, $N=11$) immersed in a bath of passive particles with packing fraction $\phi_p=0.03$. The video shows that the star polymer efficiently captures and transports passive particles as a consequence of the \textit{active embracement} mechanism. \\

\noindent
\textbf{Supplementary Video 4:} The two panels show, respectively, simulations of an active and a passive star polymer. In both cases, a single star polymer ($f=3$, $N=11$) is immersed in a bath of passive particles with packing fraction $\phi_p=0.2$. Owing to the \textit{active embracement} mechanism, the active star polymer tends to capture and transport passive particles. In contrast, the passive star polymer does not exhibit this behavior, and passive particles are instead excluded from its vicinity by steric interactions.\\

\noindent
\textbf{Supplementary Video 5:} The three panels show simulations of an active star polymer with different frictional interactions between the particles. From left to right, the simulations include: no friction, contact friction only, and both contact and rolling friction. In the absence of friction, the star polymer arms predominantly repel one another and remain in an extended configuration. Introducing contact friction suppresses this behavior and promotes the formation of collapsed configurations. When both contact and rolling friction are included, the star polymer spends most of its time in a collapsed state, in which the arms wrap around themselves and around one another.

\par

    \bibliographystyle{apsrev4-1}

    \bibliography{biblio.bib}

\section{Acknowledgments} 

\subsection{Funding}
LC acknowledges funding from the Italian Ministero dell’Università e della Ricerca under the programme PRIN 2022 ("re-ranking of the final lists"), number 2022KWTEB7, cup B53C24006470006.
HL acknowledges support by the Deutsche Forschungsgemeinschaft (DFG) through the SPP 2265, under grant numbers LO418/25. DB acknowledges the computational resources offered by CINECA in the frame of project INF26\_biophys. IA acknowledges support by the Deutsche Forschungsgemeinschaft (DFG) under project number 556762905 — AB 1083/1-1.

\subsection{Author contributions}
MM did the experiment and performed the experimental data analysis.
DB performed numerical simulations and analyzed numerical data. MM, DB, IA and LC wrote the first draft of the paper, while LC, HL and LT conceived the project and supervise experimental and numerical work.
All authors discussed the results and contributed to writing the manuscript.\par

\subsection{Competing interests}
The authors declare no competing interests.\par

\subsection{Data availability}
The experimental and simulation data generated in this study have been deposited in the Zenodo public repository at the link \cite{musacchio2026data}. 

\end{document}